\documentclass[a4paper,aps,twocolumn]{revtex4}

\usepackage{amsmath}
\usepackage{amssymb}  % for \therefore
\usepackage{graphicx}
\usepackage{mathptmx}

\newcommand{\dee}{\textnormal{d}}
\newcommand{\erf}{\,\text{erf}}

\begin{document}

\title{Equilibrium probability distribution of a conductive sphere's floating charge in a collisionless, drifting Maxwellian plasma}
\date{\today}
\author{Drew M.\ Thomas}
\email{dmt107@imperial.ac.uk}
\author{Michael Coppins}
\email{m.coppins@imperial.ac.uk}
\affiliation{Blackett Laboratory, Imperial College London, Prince Consort Road, London, SW7 2BW, UK}

\begin{abstract}
A dust grain in a plasma has a fluctuating electric charge, and past work concludes that spherical grains in a stationary, collisionless plasma have an essentially Gaussian charge probability distribution.
This paper extends that work to flowing plasmas and arbitrarily large spheres, deriving analytic charge probability distributions up to normalizing constants. We find that these distributions also have good Gaussian approximations, with analytic expressions for their mean and variance.
\end{abstract}


\maketitle

\section{Introduction}

A dust grain in a plasma acquires an electric charge by collecting electrons and ions that land on it. Because electrons and ions arrive at the grain at random, the grain's charge fluctuates, and because this fluctuating charge affects the dust's physical behaviour, the probability distribution of the charge is of practical interest.

Deriving this probability distribution for an arbitrary grain away from equilibrium in a plasma of arbitrary collisionality is very difficult, so we consider here a simpler case: a spherical, conductive grain at equilibrium in a collisionless plasma. This problem has been solved before \cite{Draine87,Matsoukas95,Matsoukas96,Zagorodny00} but these solutions have two major limitations. Firstly, they assume a stationary plasma. Secondly, most of these solutions apply the OML (orbital motion limited) grain charging theory, but OML is limited by its requirement that the grain be small relative to the Debye length \cite{Willis10}.

In this paper we go beyond previous work by relaxing these assumptions. Instead of OML we use the more general charging model SOML (shifted OML), which allows for plasma flow by assuming a shifted Maxwellian ion velocity distribution far from the sphere \cite{Hutch05,Willis11,Willis12}.
We also circumvent the small sphere requirement that orthodox SOML inherits from OML \cite[pp.\ 94--95]{Willis11}, by applying SOML differently for arbitarily large spheres.
Ultimately we derive two sets of results which, between them, cover spheres of all sizes (except those so small that their average charge is of order $e$).
One set of results applies to ``mid-sized'' spheres, the other to ``large'' spheres. A ``mid-sized'' sphere is one with $a \lesssim \lambda_D$ and a ``large'' sphere is one with $a \gg \lambda_D$, where $a$ is the sphere's radius and $\lambda_D$ the Debye length. We use these definitions of ``mid-sized'' and ``large'' throughout the paper.

We start by building two stochastic models of sphere charging, one based on ordinary SOML and the other on the alternative large sphere model. We then solve both stochastic models for their equilibrium probability distributions, and derive Gaussian approximations to both.

\section{Electron and ion currents}

Before building our stochastic charging models we need the electron and ion currents onto a sphere. In this paper we assume the electrons conform to the Boltzmann relation, and hence that a sphere of radius $a$ and surface electric potential $\phi$ has an electron current
\begin{equation}
\label{eq:Ie}
I_e = -4 \pi a^2 n_0 e \sqrt{\frac{k_B T_e}{2\pi m_e}} \exp \left( \frac{e \phi}{k_B T_e} \right)
\end{equation}
where $n_0$ is the electron density far from the sphere and $T_e$ the electron temperature. The Boltzmann relation works well regardless of plasma flow because flow is invariably much slower than the electron thermal speed.

To assume the Boltzmann relation we require $\phi < 0$. This is sometimes false for spheres so tiny that random fluctuations can render their charge positive, so we exclude those spheres from our calculations. This leaves us with mid-sized ($a \lesssim \lambda_D$) and large ($a \gg \lambda_D$) spheres.

SOML gives good estimates for ion currents onto the former.
According to SOML \cite{Shull78,Willis12}, the ion current is
\begin{equation}
\label{eq:somlIi}
I_i = 4 \pi a^2 n_0 Z e \sqrt{\frac{k_B T_i}{2\pi m_i}} \left( s_1(v) - s_2(v) \frac{Ze \phi}{k_B T_i} \right)
\end{equation}
when $\phi \le 0$, where $Z$ is the ions' charge state, $T_i$ is the ion temperature, $m_i$ is the ion mass, $v$ is the flow velocity normalized by $\sqrt{2 k_B T_i / m_i}$, and
\begin{align}
s_1(v) & \equiv \sqrt{\pi} \frac{(1 + 2v^2) \erf(v)}{4v} + \frac{\exp(-v^2)}{2} \\
s_2(v) & \equiv \sqrt{\pi} \frac{\erf(v)}{2v}
\end{align}
are auxiliary functions of the normalized flow velocity. Both functions tend to 1 in the no flow limit ($v \rightarrow 0$).
% Interesting to note that if I re-derive s_2(v) myself from Shull myself, I get a 4v in the denominator instead of 2v. Can't see why, and 2v is plainly the right choice because only then does s_2(0) = 1. Did Shull have a typo that Willis et al. fixed? What's up with this?
SOML also gives $I_i$ for $\phi > 0$ but the resulting stochastic model is intractably complicated --- see appendix \ref{S:intractability}.

For large spheres the ion current is less than that obtained from eq.\ \ref{eq:somlIi} because a larger $a/\lambda_D$ brings into existence absorption radii that undermine the SOML model \cite{Willis12}.
To avoid this problem we follow Willis et al.\ in assuming that these absorption radii are ``at or within the sheath'', and therefore apply SOML at the sheath edge instead of the sphere's surface \cite{Willis12}. $I_i$ is then given by equation \ref{eq:somlIi}, but with the sheath edge potential $\phi_s$ substituting for $\phi$. To find $\phi_s$, we suppose that ions cross the sheath edge perpendicularly at the Bohm speed $u_B$ (irrespective of flow) and make the thin sheath ($a \gg \lambda_D$) assumption that all ions crossing the sheath hit the sphere. This implies a second expression for $I_i$,
\begin{equation}
\label{eq:largeIi}
I_i = 4 \pi a^2 n_0 Z e u_B \, \exp \left( \frac{e \phi_s}{k_B T_e} \right)
\end{equation}
where the ion density at the sheath edge follows from assuming quasineutrality at the sheath edge. Equating this with eq.\ \ref{eq:somlIi} and setting $\phi = \phi_s$ gives
\begin{equation}
\label{eq:Gin01}
u_B \exp \left( \frac{e \phi_s}{k_B T_e} \right)
= \sqrt{\frac{k_B T_i}{2\pi m_i}} \left( s_1(v) - s_2(v) \frac{Ze \phi_s}{k_B T_i} \right)
\end{equation}
Estimating the Bohm speed as
\begin{equation}
u_B = \sqrt{\frac{k_B (T_e + \gamma T_i)}{m_i}}
\end{equation}
where $\gamma$ is the heat capacity ratio \cite{Willis12}, and substituting it into eq.\ \ref{eq:Gin01} gives
\begin{equation}
\sqrt{\gamma + \frac{1}{\Theta}} \,
\exp \left( \frac{e \phi_s}{k_B T_e} \right)
= \frac{ \displaystyle s_1(v) - s_2(v) \frac{Ze \phi_s}{k_B T_i} }{\sqrt{2\pi}}
\end{equation}
where $\Theta \equiv T_i / T_e$ is the ion-to-electron temperature ratio. Solving this equation for $\phi_s$ gives
\begin{equation}
\label{eq:phis}
\phi_s = \frac{k_B T_e}{e} \left(c - W \left( \frac{\sqrt{2\pi} \, c}{s_1(v)} \sqrt{\gamma + \frac{1}{\Theta}} \exp(c) \right) \right)
\end{equation}
where
\begin{equation}
\label{eq:c}
c \equiv \frac{\Theta s_1(v)}{Z s_2(v)}
\end{equation}
and $W(x)$ is the principal branch of the Lambert W special function \cite{Dubinov05}. The expression for $\phi_s$ is unwieldy but has the nice property of being independent of the sphere's charge.

We assume that the plasma contains only one species of singly charged positive ion, so we can simply take $Z = +1$ and $c = \Theta s_1(v) / s_2(v)$ from this point on.

\section{Building the stochastic models}

Over a sufficiently short period of time $\delta$, it is extremely unlikely that the sphere has time to collect multiple particles. Hence during a small enough $\delta$ effectively only three events may happen: the sphere absorbs nothing; the sphere absorbs an electron; or the sphere absorbs an ion. The chances of these events happening during a given $\delta$ depend (to a good approximation) on only the sphere's net charge $Ne$ at the start of that period. As such we can model the sphere's charge fluctuations as a Markovian ``one-step process'' \cite[p.\ 134]{vanKampen92}, where the sphere's state is its net charge (in elementary charges) $N$, and $N$ changes only in sporadic increments of $\pm 1$.

A one-step process is characterized by its rate coefficients $r_N$, the probability per unit time of a shift from state $N$ to state $N-1$, and $g_N$, the probability per unit time of a shift from state $N$ to state $N+1$ \cite[p.\ 134]{vanKampen92}. In our model these rates correspond to the electron collection rate $\dot{N}_e$ and ion collection rate $\dot{N}_i$.

The electron collection rate is
\begin{equation}
\dot{N}_e = \frac{I_e}{-e} = \chi \exp \left( \frac{e \phi}{k_B T_e} \right)
\end{equation}
where
\begin{equation}
\chi \equiv 4 \pi a^2 n_0 \sqrt{\frac{k_B T_e}{2\pi m_e}}
\end{equation}
is the collection rate for a neutral sphere of the same size.

The ion collection rate depends on which potential we use in equation \ref{eq:somlIi}. For $a \lesssim \lambda_D$ we use the surface potential $\phi$, but for $a \gg \lambda_D$ we use the sheath edge potential $\phi_s$, as explained above. To accommodate both options we write
\begin{equation}
\label{eq:Nidot}
\dot{N}_i = \frac{I_i}{e} = \chi \frac{\sqrt{\Theta}}{\mu} \left( s_1(v) - s_2(v) \frac{e \phi'}{k_B T_i} \right)
\end{equation}
where $\mu \equiv \sqrt{m_i / m_e}$ is a normalized ion mass, and $\phi'$ is either $\phi$ (if $a \lesssim \lambda_D$) or $\phi_s$ (if $a \gg \lambda_D$). This will lead to two different models, one for mid-sized grains and one for large grains.

For both models the rate coefficients $r_N$ and $g_N$ are
\begin{equation}
\label{eq:rNdef}
r_N = \dot{N}_e \delta = \chi \delta \exp \left( \frac{e \phi}{k_B T_e} \right)
\end{equation}
and
\begin{equation}
\label{eq:gNdef}
g_N = \dot{N}_i \delta = \chi \delta \frac{\sqrt{\Theta}}{\mu} \left( s_1(v) - s_2(v) \frac{e \phi'}{k_B T_i} \right) \\
\end{equation}

To complete the stochastic models, one must define $\phi$ in terms of $N$. Assuming the sphere is conducting, and neglecting polarization from nearby plasma particles, $\phi$ is
\begin{equation}
\label{eq:phiN}
\phi = \frac{e}{4 \pi \varepsilon_0 a} N = \frac{k_B T_e \alpha}{e} N
\end{equation}
where
\begin{equation}
\alpha \equiv \frac{e^2}{4 \pi \varepsilon_0 k_B T_e a}
\end{equation}
is a dimensionless characteristic parameter analogous to the Coulomb coupling parameter $\Gamma_c$ for a simple plasma; in fact $\alpha = (d / a) \Gamma_c$, where $d$ is the average inter-particle distance.
The parameter $\alpha$ is key to the results that follow, with many of our approximations relying on its being small ($\alpha \ll 1$ or $\alpha \ll \Theta \lesssim 1$).

Substituting eq.\ \ref{eq:phiN} into equation \ref{eq:rNdef},
\begin{equation}
r_N = \chi \delta \exp \left( \alpha N \right)
\label{eq:rN}
\end{equation}
For $a \lesssim \lambda_D$ we likewise substitute eq.\ \ref{eq:phiN} for $\phi'$ in eq.\ \ref{eq:gNdef}:
\begin{equation}
g_N = \chi \delta \frac{\sqrt{\Theta}}{\mu} \left( s_1(v) - \frac{\alpha s_2(v)}{\Theta} N \right)
\label{eq:gN1}
\end{equation}
which we rewrite as
\begin{equation}
\label{eq:gN2}
g_N = \chi \delta \frac{\alpha s_2(v)}{\mu \sqrt{\Theta}} \left( \frac{c}{\alpha} - N \right)
\end{equation}
to streamline the next section's algebra. For $a \gg \lambda_D$, $\phi = \phi_s$ and so eq.\ \ref{eq:gNdef} is independent of $N$, being
\begin{equation}
\label{eq:gN3}
g_N = g \equiv \chi \delta \frac{s_2(v)}{\mu \sqrt{\Theta}}
W \left( \frac{\sqrt{2 \pi} \, c}{s_1(v)} \sqrt{\gamma + \frac{1}{\Theta}} \exp(c) \right)
\end{equation}
where we drop the $N$ subscript to emphasize the independence from $N$ for large spheres.

We now have one complete stochastic model for mid-sized grains (comprising eqs.\ \ref{eq:rN} \& \ref{eq:gN2}) and one for large grains (comprising eqs.\ \ref{eq:rN} \& \ref{eq:gN3}).
The next step is solving them for the charge probability distribution $f_N$. Normally one would solve each model's master equation \cite[passim]{vanKampen92}, but these models' master equations aren't exactly solvable. For an exact solution we use a more direct approach.

\section{Solving the stochastic models}

\label{S:solving}

At equilibrium detailed balance holds \cite[p.\ 142]{vanKampen92}. That is, in a huge ensemble of sphere-in-plasma systems at equilibrium, just as many should be going from charge state $N$ to $N-1$ as are going from charge state $N-1$ to $N$ (on average). As such
\begin{equation}
r_N f_N = g_{N-1} f_{N-1}
\label{eq:detailed}
\end{equation}
which is a recurrence relation that has $f_N$ as its solution.

We solve it for the $a \lesssim \lambda_D$ case first. Substituting eqs.\ \ref{eq:rN} and \ref{eq:gN2} into eq.\ \ref{eq:detailed},
\begin{equation}
\label{eq:rec1}
\frac{f_{N-1}}{f_N} = \frac{\mu\sqrt{\Theta}}{\alpha s_2(v)} \frac{\exp(\alpha N)}{\frac{c}{\alpha} + 1 - N}
\end{equation}
for $N \le 0$.
%Taking this ratio cancels out $\chi$ and $\delta$.
%To solve eq.\ \ref{eq:rec1}, note that
To solve this equation, note that
\begin{align}
\label{eq:fprodform}
f_N = & \ f_0 \prod_{M=N+1}^0 \frac{\mu\sqrt{\Theta}}{\alpha s_2(v)} \exp(\alpha M) \\
& \div \prod_{M=N+1}^0 {\frac{c}{\alpha} + 1 - M} \notag
\end{align}
for $N < 0$.
By inspection the first product is
\begin{equation}
\left(\frac{\alpha s_2(v)}{\mu\sqrt{\Theta}} \right)^N
\exp \left(- \frac{\alpha}{2} N(N+1) \right)
\label{eq:firstprod}
\end{equation}
and the second is equivalent to
\begin{equation}
\left( \prod_{M=0}^{-N-1} 1 + \frac{c}{\alpha} + M \right) = \left( 1 + \frac{c}{\alpha} \right)^+_{-N}
\label{eq:pochhammer1}
\end{equation}
where $(x)^+_n \equiv $ is the rising factorial (or ``Pochhammer symbol''), defined as
\begin{equation}
(x)^+_n \equiv x (x+1) (x+2) \cdots (x+n-1)
\end{equation}
%The Pochhammer symbol can be defined as a ratio of gamma functions:
%\begin{equation}
%(z)^+_n \equiv \frac{\Gamma(z+n)}{\Gamma(z)}
%\label{eq:pochgamma}
%\end{equation}
%which implies that
%\begin{align}
%(N+c)^+_{-N}
%& \equiv \left( \frac{\Gamma(N+c)}{\Gamma(N+c-N)} \right)^{-1} \\
%& \equiv \left( \frac{\Gamma(N+c)}{\Gamma(c)} \right)^{-1} \notag \\
%& \equiv 1 / (c)^+_N \notag
%\end{align}
%This allows me to remove $N$ from the interior of the Pochhammer symbol in eq.\ \ref{eq:pochhammer1}, rewriting that expression as
%\begin{equation}
%\left( - \frac{\Theta}{\alpha s_2(v)} \right)^N \div (c)^+_N
%\label{eq:pochhammer2}
%\end{equation}
%where
%\begin{equation}
%c \equiv - \frac{1}{\alpha} \left( \frac{e}{k_B T_e} \Delta \phi + \Theta \frac{s_1(v)}{s_2(v)} \right)
%\label{eq:pochcdef}
%\end{equation}
Putting together eqs.\ \ref{eq:fprodform}, \ref{eq:firstprod} \& \ref{eq:pochhammer1},
\begin{align}
\label{eq:f1}
f_N = & \ f_0
\left( \frac{\alpha s_2(v)}{\mu \sqrt{\Theta}} \right)^N \\
& \times \exp \left( - \frac{\alpha}{2} N(N+1) \right)
\div \left( 1 + \frac{c}{\alpha} \right)^+_{-N} \notag
\end{align}
where $f_0$ is determined by the normalization condition
\begin{equation}
\sum_{N=-\infty}^{0} f_N = 1
\end{equation}
which is, unfortunately, not analytically solvable. However, equation \ref{eq:f1} permits numerical calculation of $N$'s probability distribution in practice; one can simply compute $\sum f_N / f_0$ for those $N$ where $f_N/f_0$ is non-negligible, and set equation \ref{eq:f1}'s $f_0$ to that sum's reciprocal.

As one may rewrite $(x)^+_n$ as a ratio of gamma functions (except when $x$ or $x+n$ is a negative integer), eq.\ \ref{eq:f1} constitutes an analytic definition of $f_N$ in terms of elementary functions and the gamma function, lacking only $f_0$'s value:
\begin{align}
f_N = & \ f_0
%\left( \frac{\alpha s_2(v)}{\mu \sqrt{\Theta}} \right)^N \exp \left( - \alpha \frac{N(N+1)}{2} \right) \\
\exp \left( N \ln \left( \frac{\alpha s_2(v)}{\mu \sqrt{\Theta}} \right)
- \frac{\alpha}{2} N(N+1) \right) \\
& \times \frac{\Gamma \left( 1 + c/\alpha \right)}{\Gamma (1 + c/\alpha - N)} \notag
\end{align}

We now turn to the $a \gg \lambda_D$ case.
Substituting eqs.\ \ref{eq:rN} and \ref{eq:gN3} into eq.\ \ref{eq:detailed},
\begin{equation}
\label{eq:fnfnlarge}
\frac{f_{N-1}}{f_N} = \frac{\exp(\alpha N)}{g^*}
\end{equation}
where
%$g^* \equiv \frac{g}{\chi \delta}$
$g^* \equiv g / (\chi \delta)$
is a more convenient form of $g$. The implied product of exponentials is readily solvable by inspection:
\begin{equation}
f_N = f_0 {g^*}^N \exp \left( - \frac{\alpha}{2} N (N+1) \right)
\end{equation}
The normalization condition does not appear to be analytically solvable for this distribution, either.

\section{The modal charge and the stochastic model's validity}

Deriving a closed form expression for $f_N$'s mode is conceptually straightforward.
Given the mode at $M$, $f_{N-1}/f_N \le 1$ for $N < M$, and $f_{N-1}/f_N \ge 1$ for $N > M$.
Therefore, because $f_{N-1}/f_N$ is monotonic (q.v.\ appendix \ref{S:unimod}), the mode $M$ is located where $f_{M-1} / f_M \approx 1$.
(It is safe to refer to ``the'' mode because $f_N$ is unimodal, as we show in appendix \ref{S:unimod}.)

We derive $M$ for mid-sized grains first. Setting equation \ref{eq:rec1}'s left hand side to 1, substituting $M$ for $N$, and solving,
\begin{equation}
\label{eq:themode}
M \approx 1 + \frac{\displaystyle c - W \left( \frac{\mu\sqrt{\Theta}}{s_2(v)} \exp(\alpha + c) \right)}{\alpha}
\end{equation}
where $W(x)$ is again the Lambert W special function's principal branch.
For small $\alpha$ (i.e.\ large $T_e a$), $M$'s dependence on the Lambert W term is weak and $M$'s dependence on $\alpha$ goes approximately as $\mathcal{O}(1/\alpha)$.

Equation \ref{eq:themode} leads to an obvious precondition for the stochastic model's validity. As the model assumes the sphere never has a positive charge, a positive value of $M$ indicates that the model has broken down and become self-inconsistent. Therefore $M \le 0$ is a necessary condition for model validity.

When is $M \le 0$? Rearranging eq.\ \ref{eq:themode}, $M \le 0$ when
\begin{equation}
\alpha \lesssim W \left( \frac{\mu\sqrt{\Theta}}{s_2(v)} \exp(\alpha + c) \right) - c
\end{equation}
By exploiting $W(z)$'s definition, monotonicity and positivity for positive arguments, one can rewrite the inequality to remove the right hand side's dependence on $\alpha$:
\begin{equation}
\label{eq:Mineq1}
\alpha \lesssim \frac{\mu \sqrt{\Theta}}{s_2(v)} - c
\end{equation}
This sets an upper bound on $\alpha$, above which the inequality is unsatisfied and the model fails. This is not surprising, as the model assumes the sphere is not very tiny, which implies a small $\alpha \propto 1/(T_e a)$.

Substituting in $c$ (eq.\ \ref{eq:c}),
\begin{equation}
\label{eq:Mineq2}
\alpha \lesssim \frac{\mu \sqrt{\Theta} - \Theta s_1(v)}{s_2(v)}
\end{equation}
Evidently, in the cold ion limit ($\Theta \rightarrow 0$), the inequality reduces to $\alpha \lesssim 0$ and is never satisfied, indicating model failure. This is also unsurprising, as a vanishing $\Theta$ requires either an infinite $T_e$ (and hence an infinite electron current, from eq.\ \ref{eq:Ie}) or a zero $T_i$ (and hence an infinite ion current whenever $\phi < 0$).
%Moreover, orbital-motion limited models apply where $T_i$ is significant; where it is insignificant a radial motion charging model is more appropriate \cite{KA02}.

Inversely, for vanishing $T_e$ ($\Theta \rightarrow +\infty$), the inequality's right hand side tends to $-\infty$, in which case the inequality is again never fulfilled and the model fails.

\begin{figure}
\includegraphics[width=0.49\textwidth]{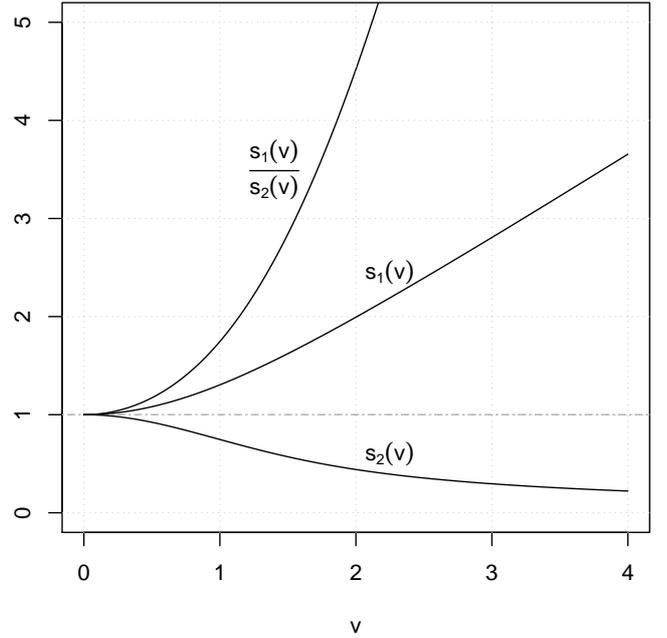}
\caption{\label{fig:s1s2} The auxiliary functions $s_1(v)$ and $s_2(v)$, and their ratio, as a function of $v$ (the plasma drift velocity normalized by $\sqrt{2 k_B T_i / m_i}$).}
\end{figure}

The inequality also shows that flow affects the model's validity.
Because $\Theta > 0$ and $s_1(v)$ increases with $v$ (figure \ref{fig:s1s2}), equation \ref{eq:Mineq2}'s right hand side becomes negative for large $v$, and the model eventually fails. Fortunately this only occurs at truly huge flow velocities, as shown by solving the trivial sub-inequality
\begin{equation}
\mu \sqrt{\Theta} - \Theta s_1(v) < 0
\quad \Rightarrow s_1(v) > \frac{\mu}{\sqrt{\Theta}}
\end{equation}
Because ions are rarely hotter than electrons, and the lightest ions are protons, $\mu/\sqrt{\Theta} \gtrsim \sqrt{m_p/m_e} = 42.85$, which $s_1(v)$ is always less than for any reasonable flow speed ($v < 48.34$).
%From figure \ref{fig:s1s2}, $s_1(v)$ is only greater than 42.85 for exceptionally fast flows.

More sedate flow velocities have the effect of \emph{increasing} eq.\ \ref{eq:Mineq2}'s right hand side, as shown by calculating that
\begin{align}
\frac{1}{s_2(v)} & \ = \ 1 + \frac{1}{3} v^2 + \mathcal{O} \left( v^4 \right) \\
\frac{s_1(v)}{s_2(v)} & \ = \ 1 + \frac{2}{3} v^2 + \mathcal{O} \left( v^4 \right) & 
\end{align}
and substituting into equation \ref{eq:Mineq2}:
\begin{equation}
\alpha \lesssim \left( \mu \sqrt{\Theta} - \Theta \right) + \frac{1}{3} \left( \mu \sqrt{\Theta} - 2 \Theta \right) v^2
\end{equation}
Inevitably $\mu \sqrt{\Theta} > 2\Theta$, so flow's effect (to second order) is to increase the right hand side, raising the chance of satisfying the inequality and loosening the validity constraint. The second order effect can also compensate for a low $\Theta$; as $\Theta$ shrinks, $\mu \sqrt{\Theta}$ gets greater relative to the negative $\mathcal{O}(\Theta)$ terms, enhancing flow's beneficial effect on the model's validity.

Deriving $M$ for large grains is trivial. Setting equation \ref{eq:fnfnlarge}'s left hand side to 1 immediately leads to
\begin{equation}
\label{eq:themodelarge}
M \approx \frac{\ln g^*}{\alpha}
\end{equation}
The direct $1/\alpha$ dependence dovetails with eq.\ \ref{eq:themode}'s approximate $1/\alpha$ dependence for small $\alpha$.

The same $M \le 0$ validity condition applies here. From eq.\ \ref{eq:themodelarge}, $M \le 0$ if and only if $g^* \le 1$, i.e.\ if
\begin{equation}
\label{eq:largvalid}
\frac{s_2(v)}{\sqrt{\Theta}}
W \left( \frac{\sqrt{2 \pi} \, \Theta}{s_2(v)} \sqrt{\gamma + \frac{1}{\Theta}} \exp(c) \right)
\lesssim \mu
\end{equation}
We can get some insight from this knotty expression by considering limiting cases. For example, we can expand about $\Theta = 0$ to obtain a validity inequality for a cold ion plasma:
\begin{equation}
\sqrt{2\pi} - \frac{2\pi \sqrt{\Theta}}{s_2(v)} + \mathcal{O}(\Theta) \lesssim \mu
\end{equation}
Neglecting higher order terms, this inequality is always true, since the LHS is always less than $\sqrt{2\pi}$, and $\sqrt{2\pi} < \mu$. With a large grain and cold ions the stochastic model is always self-consistent, regardless of flow.

Another limiting case is that of vanishing flow velocity. With $v=0$, $s_1(v) = s_2(v) = 1$, and eq.\ \ref{eq:largvalid} becomes
\begin{equation}
W \left( \sqrt{2 \pi \left( \gamma + \frac{1}{\Theta} \right)} \, \Theta \, \exp(\Theta) \right)
\lesssim \mu \sqrt{\Theta}
\end{equation}
By exploiting $W(z)$'s definition, monotonicity and positivity for positive arguments once again,
\begin{equation}
\label{eq:largvalid2}
\sqrt{2 \pi \left( \gamma \Theta + 1 \right)}
\exp(\Theta)
\lesssim
\mu
\exp(\mu \sqrt{\Theta})
\end{equation}
For ease of solution, we replace this with a more stringent validity condition. Specifically, when
\begin{equation}
\label{eq:largvalid3}
\sqrt{2 \pi (\gamma \Theta + 1)} < \mu
\end{equation}
the inequality
\begin{equation}
\exp(\Theta) \lesssim \exp(\mu \sqrt{\Theta})
\end{equation}
is clearly an even tighter bound on $\Theta$ than eq.\ \ref{eq:largvalid2}. This tighter condition quickly reduces to $\Theta \lesssim \mu^2$. This bound, together with eq.\ \ref{eq:largvalid3}, implies the condition
\begin{equation}
\Theta \lesssim \frac{1}{\gamma} \left( \frac{\mu^2}{2\pi} - 1 \right)
\end{equation}
when $\gamma \ge 1/(2\pi)$, which is always true because $\gamma$ is between 1 and 3 \cite{Willis10}. Even when $\mu^2$ is as small as realistically possible ($m_p / m_e = 1836$) and $\gamma$ as large as realistically possible (3), $\Theta$ must be ludicrously high (at least 97) to violate even this conservative validity condition.

Thus the large grain model is always valid when at least one of $\Theta$ or $v$ is small. To find a regime where the model breaks down, we now consider the large $v$ limit. For large $v$,
\begin{align}
s_2(v) & \rightarrow \frac{\sqrt{\pi}}{2v} \\ 
\frac{s_1(v)}{s_2(v)} & = v^2 + \frac{1}{2} + \frac{v \exp(-v^2)}{\sqrt{\pi} \, \erf(v)} \rightarrow v^2 + \frac{1}{2}
\end{align}
and eq.\ \ref{eq:largvalid} becomes
\begin{equation}
W \left( 2 \sqrt{2} \, \Theta v \sqrt{\gamma + \frac{1}{\Theta}} \exp \left( \Theta v^2 + \frac{\Theta}{2} \right) \right)
\lesssim \frac{2 \mu \sqrt{\Theta} \, v}{\sqrt{\pi}}
\end{equation}
Taking the inverse Lambert W function of both sides and cancelling common terms,
\begin{equation}
%\frac{\sqrt{2 \pi} \, \Theta}{s_2(v)} \sqrt{\gamma + \frac{1}{\Theta}} \exp(c)
\sqrt{2 (\Theta \gamma + 1)} \exp \left( \Theta v^2 + \frac{\Theta}{2} \right)
\lesssim
\frac{\mu}{\sqrt{\pi}}
\exp \left( \frac{2 \mu \sqrt{\Theta} \, v}{\sqrt{\pi}} \right)
\end{equation}
Taking logarithms and rearranging,
\begin{equation}
\Theta v^2
- \frac{2 \mu \sqrt{\Theta} \, v}{\sqrt{\pi}}
\lesssim
\ln \frac{\mu}{\sqrt{2 \pi (\Theta \gamma + 1)}}
- \frac{\Theta}{2}
\end{equation}
The right hand side is smallest when $\mu$ is smallest and $\gamma$ and $\Theta$ are largest. Realistically, $\mu \ge 42.85$, $\gamma \le 3$ and $\Theta \lesssim 1$, so the RHS is at least 1.6. As such, setting the RHS to zero gives the tighter inequality
\begin{equation}
\Theta v^2
- \frac{2 \mu \sqrt{\Theta} \, v}{\sqrt{\pi}} \lesssim 0
\end{equation}
which gives the conservative velocity limit
\begin{equation}
v \lesssim \frac{2 \mu}{\sqrt{\pi \Theta}}
\end{equation}
Like the mid-sized grain model, the large grain model breaks down only in the face of exceptional flow ($v \sim 50$).

\section{The master equation and Gaussian approximations}

\label{S:masterGaussian}

Although the stochastic models' master equations have no exact, analytic solution, we can follow Matsoukas, Russell \& Smith \cite{Matsoukas94,Matsoukas95,Matsoukas96} in finding approximate solutions by treating the models as if continuous. A one-step process has the master equation \cite[p.\ 134]{vanKampen92}
\begin{align}
\frac{\partial f_N(t)}{\partial t} = & \ r_{N+1} f_{N+1}(t) + g_{N-1} f_{N-1}(t) \\
& \ - (r_N + g_N) f_N(t) \notag
\end{align}
This one-step master equation is approximated well by the following Fokker-Planck equation when $r_N$ and $g_N$ are smooth, slowly varying functions of $N$ \cite[pp.\ 197--198 \& 207--208]{vanKampen92}:
\begin{align}
\label{eq:fp}
\frac{\partial f_N(t)}{\partial t} = & \ -\frac{\partial}{\partial N} (g_N - r_N) f_N(t) \\
& + \frac{1}{2} \frac{\partial^2}{\partial N^2} (r_N + g_N) f_N(t) \notag
\end{align}
For our models, the F-P approximation's conditions are satisfied when the equilibrium $N$ is large. Except for the tiniest grains, the equilibrium $N$ is approximately $M$, which is of order $1/\alpha$ for mid-sized grains satisfying $\alpha \ll 1$ and of order $(\ln g^*) / \alpha$ for large grains. Thus the F-P approximation is a good one for mid-sized grains when $\alpha \ll 1$, and for large grains when $\alpha \ll -\ln g^*$.

At equilibrium, eq.\ \ref{eq:fp}'s left hand side is nil. This banishes $f_N(t)$'s time dependence, so we write the equilibrium probability distribution as $f_N$ as before. Integrating both sides with respect to $N$,
\begin{equation}
s = -(g_N - r_N) f_N + \frac{1}{2} \frac{\dee}{\dee N} (r_N + g_N) f_N
\end{equation}
where $s$ is a constant of integration corresponding to the relative probability current between charge states \cite[p.\ 72]{Risken84}. At equilibrium this current is a constant, and for this system must be zero because $N$ is bounded \cite[p.\ 98]{Risken84}. Applying the boundary condition $s = 0$ and then the product rule,
\begin{align}
0 = & \ (r_N - g_N) f_N \\
& + \frac{1}{2} \left( (r_N + g_N) f'_N + f_N \frac{\dee (r_N + g_N)}{\dee N} \right) \notag
\end{align}
where $f'_N \equiv \dee f / \dee N$.
Rearranging,
\begin{equation}
\frac{f'_N}{f_N} = \frac{\dee \ln f_N}{\dee N} = y(N)
\end{equation}
where
\begin{equation}
y(N) \equiv \frac{2g_N - 2r_N - \frac{\dee}{\dee N} (g_N + r_N)}{g_N + r_N}
\end{equation}
Then
\begin{equation}
\label{eq:fasyexp}
f_N = \exp \left( \int y(N) \; \dee N \right)
\end{equation}
The integral is insoluble for both models, but approximate solutions are possible by linearizing $y(N)$ about an $N$ where most of $f_N$'s probability density is concentrated. We could use the mode $M$ but the algebra is tidier if we use the value $N_0$ satisfying $y(N_0) = 0$. ($N_0$ is the continuous analogue of $M$, being where $f_N$ is maximized, from $y(N)$'s definition.) Then
\begin{equation}
y(N) \approx y(N_0) + (N - N_0) y'(N_0) = (N - N_0) y'(N_0)
\label{eq:ylinearization}
\end{equation}
Substituting into equation \ref{eq:fasyexp},
\begin{align}
f_N & \approx \exp \left( - y'_0 N_0 N + \frac{y'_0}{2} N^2 \right) \\
\therefore f_N & \propto \exp \left( - \frac{(N - N_0)^2}{2 / -y'_0} \right)
\end{align}
where $y'_0 \equiv y'(N_0)$ for brevity. This is a Gaussian probability distribution with mean $N_0$ and variance $-1/y'_0$, so the final approximate probability distribution is
\begin{equation}
f_N = \sqrt{\frac{-y'_0}{2\pi}}
\exp \left( - \frac{(N - N_0)^2}{2 / -y'_0} \right)
\end{equation}
Because $y(N_0) = 0$, the mean $N_0$ is implicitly defined by
\begin{equation}
\label{eq:N0defeq}
2(g_{N_0} - r_{N_0}) = \frac{\dee}{\dee N} \Big|_{N=N_0} (g_N + r_N)
\end{equation}

Inserting $r_N$ and $g_N$ for mid-sized grains (eqs.\ \ref{eq:rN} \& \ref{eq:gN2}),
\begin{equation}
\frac{\alpha s_2(v)}{\mu \sqrt{\Theta}} \left( 1 + 2 \left( \frac{c}{\alpha} - N_0 \right) \right) = (2+\alpha) \exp (\alpha N_0)
\end{equation}
This has the solution
\begin{equation}
\label{eq:gaussN0}
N_0 = \frac{1}{2} + \frac{ \displaystyle c - W \left( \frac{\mu \sqrt{\Theta}}{s_2(v)} \left(1 + \frac{\alpha}{2} \right)
\exp \left( \frac{\alpha}{2} + c \right) \right) }{\alpha}
\end{equation}
which is of similar form to the expression for $M$ (eq.\ \ref{eq:themode}), and again of order $1/\alpha$ for small $\alpha$. For a Gaussian distribution the mean equals the mode, so it makes sense that $N_0 \approx M$ algebraically.

Substituting the large grain $r_N$ and $g_N$ (eqs.\ \ref{eq:rN} \& \ref{eq:gN3}) into eq.\ \ref{eq:N0defeq} gives
\begin{equation}
2 g^* = (2 + \alpha) \exp(\alpha N_0)
\end{equation}
which has the solution
\begin{equation}
\label{eq:betterN0large}
N_0 = \frac{1}{\alpha} \ln \frac{2 g^*}{2 + \alpha}
\end{equation}
similar to the large grain formula for $M$ (eq.\ \ref{eq:themodelarge}).

Even simpler expressions for $N_0$ arise when the derivatives in eq.\ \ref{eq:N0defeq} are small compared to $g_{N_0}$ and $r_{N_0}$, which occurs when $\alpha \ll \Theta \lesssim 1$. In that regime eq.\ \ref{eq:N0defeq} reduces to
\begin{equation}
\label{eq:Gapprequate}
2(g_{N_0} - r_{N_0}) \approx 0
\end{equation}
giving the solutions
\begin{equation}
\label{eq:approxN0}
N_0 \approx \frac{1}{\alpha} \left( c - W \left( \frac{\mu \sqrt{\Theta}}{s_2(v)} \exp(c) \right) \right)
\end{equation}
for mid-sized grains (becoming close to eq.\ \ref{eq:themode} for very small $\alpha$) and
\begin{equation}
\label{eq:apprN0large}
N_0 \approx \frac{\ln g^*}{\alpha}
\end{equation}
for large grains, which matches $M$ (eq.\ \ref{eq:themodelarge}) and makes the asymptotic $\mathcal{O}(1/\alpha)$ dependence very explicit. Eq.\ \ref{eq:Gapprequate} amounts to equating $I_i$ and $I_e$, so eq.\ \ref{eq:approxN0} implies the same normalized electric potential as the SOML equation \cite[eq.\ 15]{Willis12}, which comes from explicitly taking $I_i = I_e$. The more exact $N_0$ given by eq.\ \ref{eq:gaussN0} implies a more negative electric potential.

The same simplification allows a concise approximation for $y'_0$ and so the distribution's variance. Neglecting derivatives,
\begin{equation}
y(N) \approx \frac{2 (g_N - r_N)}{g_N + r_N}
\end{equation}
and so, applying the quotient rule and simplifying,
\begin{equation}
y'(N) \approx \frac{4}{(g_N + r_N)^2}
\left( r_N \frac{\dee g_N}{\dee N} - g_N \frac{\dee r_N}{\dee N} \right)
\end{equation}

Solving for $y'_0$ for mid-sized grains is tedious but feasible. Substituting in $r_N$ and $g_N$, and applying eq.\ \ref{eq:approxN0} eventually gives
\begin{align}
\label{eq:y0prime}
y'_0 & \approx -\alpha \left( 1 + \frac{1}{W \left( \frac{\mu \sqrt{\Theta}}{s_2(v)} \exp(c) \right) } \right) \\
\therefore y'_0 & \approx -\alpha \left( 1 + \frac{1}{c - \alpha N_0} \right)
\end{align}
The variance is then
\begin{equation}
\label{eq:varmid}
\sigma^2 \approx \frac{1}{\alpha} \left( 1 + \frac{1}{c - \alpha N_0} \right)^{-1}
\end{equation}
for $\alpha \ll \Theta$, so $\sigma^2 \propto 1/\alpha \propto T_e a$ in this regime, consistent with Matsoukas and Russell's finding that $\sigma^2 \propto a$ \cite[p.\ 4288]{Matsoukas95} ($\alpha$ being proportional to $1/a$). Also consistent is the implication that the normalized standard deviation $\sigma/N_0 \propto \sqrt{\alpha} \propto 1/\sqrt{a}$.

\begin{figure}
\includegraphics[width=0.49\textwidth]{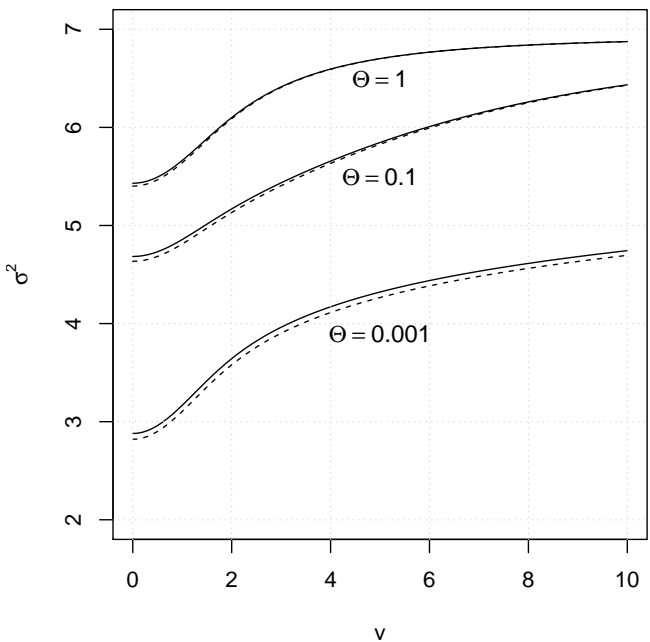}
\caption{\label{fig:fvar1} The variance $\sigma^2$ of $f_N$ for a 10 nm sphere in a hydrogenic plasma with 1 eV electrons and $a \lesssim \lambda_D$, calculated numerically from $f_N$ iself (solid lines) and from eq.\ \ref{eq:varmid} (dashed lines).}
\end{figure}

While the dependence of $\sigma^2$ on $\alpha$ goes as $\mathcal{O}(1/\alpha)$, the dependence on $v$ is more complicated. Figure \ref{fig:fvar1} shows how $\sigma^2$ nonlinearly increases with $v$. Gentle flows ($v \ll 1$) have a negligible effect on $\sigma^2$, but as the flow speed exceeds the ion thermal speed ($v \sim 1$) $\sigma^2$ rises appreciably with $v$, plateauing at a higher value for very rapid flow.
Eq.\ \ref{eq:varmid} systematically underestimates $\sigma^2$, but not appreciably.
For still faster flows the model breaks down because the sphere's chance of acquiring a positive charge becomes non-negligible.

For large grains,
\begin{equation}
y'_0 \approx \frac{-4 g}{(g + r_{N_0})^2} \frac{\dee r_N}{\dee N} \Big|_{N=N_0}
= \frac{-4 g^* \alpha \exp(\alpha N_0)}{(g^* + \exp(\alpha N_0) )^2}
\end{equation}
Applying eq.\ \ref{eq:apprN0large}, the variance is
\begin{equation}
\sigma^2 \approx 
\frac{(g^* + \exp(\alpha N_0) )^2}{4 g^* \alpha \exp(\alpha N_0)}
\approx \frac{1}{\alpha}
\approx \frac{N_0}{\ln g^*}
\end{equation}
which aligns with the asymptotic $1/\alpha$ dependence for mid-sized grains. Notice that $\sigma^2$ depends only on $\alpha$ for large grains. This remains the case even if we use the more exact value of $N_0$ given in equation \ref{eq:betterN0large}:
\begin{equation}
\sigma^2 \approx 
\frac{(g^* + \exp(\alpha N_0) )^2}{4 g^* \alpha \exp(\alpha N_0)}
= \frac{1}{\alpha} \left( 1 + \frac{\alpha^2}{16 + 8\alpha} \right)
\end{equation}
Using the more exact $N_0$ has the sole effect of making $\sigma^2$ slightly greater than $1/\alpha$; it reveals no dependence on $\mu$, $\Theta$, or $v$. We therefore reach the interesting conclusion that for a given large grain, the mean charge is sensitive to the values of the plasma parameters $\mu$, $\Theta$, $v$, and $\gamma$, but the charge's \emph{variance} is not. The variance depends only on $T_e$ (and $a$).

\section{Skewness of the charge distribution}

The Gaussian approximation to $f_N$ roughly matches $f_N$'s mean and variance, but ignores $f_N$'s higher order moments. This may be problematic if $f_N$ has appreciable skew, which is likely if it deviates a lot from a Gaussian distribution. The Gaussian approximation hinges on several assumptions, namely that the charging process is virtually continuous, with $r_N$ and $g_N$ being smooth and only weakly dependent on $N$ (so we may represent the master equation as a Fokker-Planck equation), and that $f_N$'s probability mass is concentrated around its mode (to justify the linearization embodied in eq.\ \ref{eq:ylinearization}). These assumptions never hold perfectly, so we expect a little non-Gaussianness and so a little skew. However, the skewness may be negligible for realistic parameter values.

\begin{figure}
\includegraphics[width=0.49\textwidth]{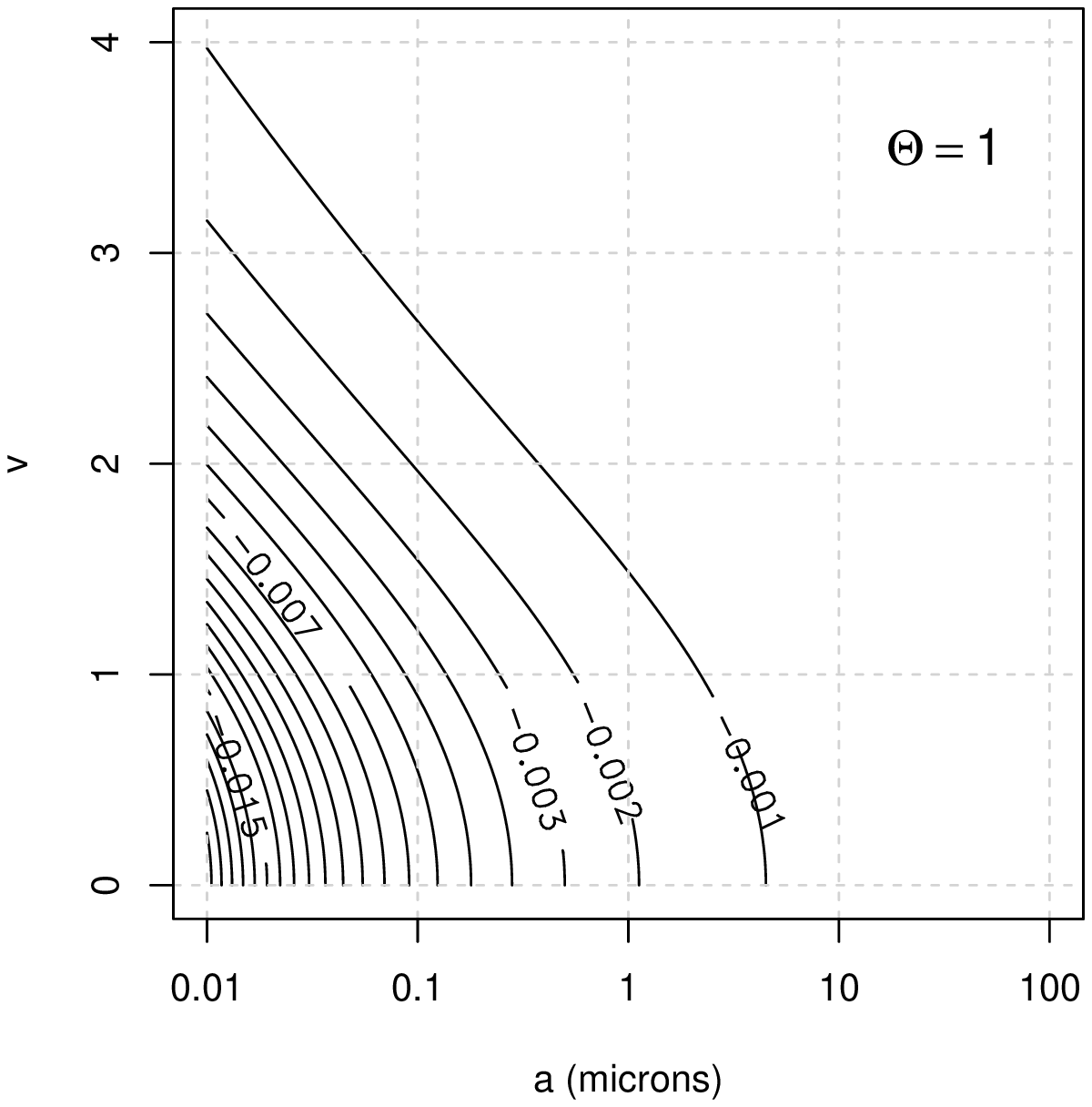}
\includegraphics[width=0.49\textwidth]{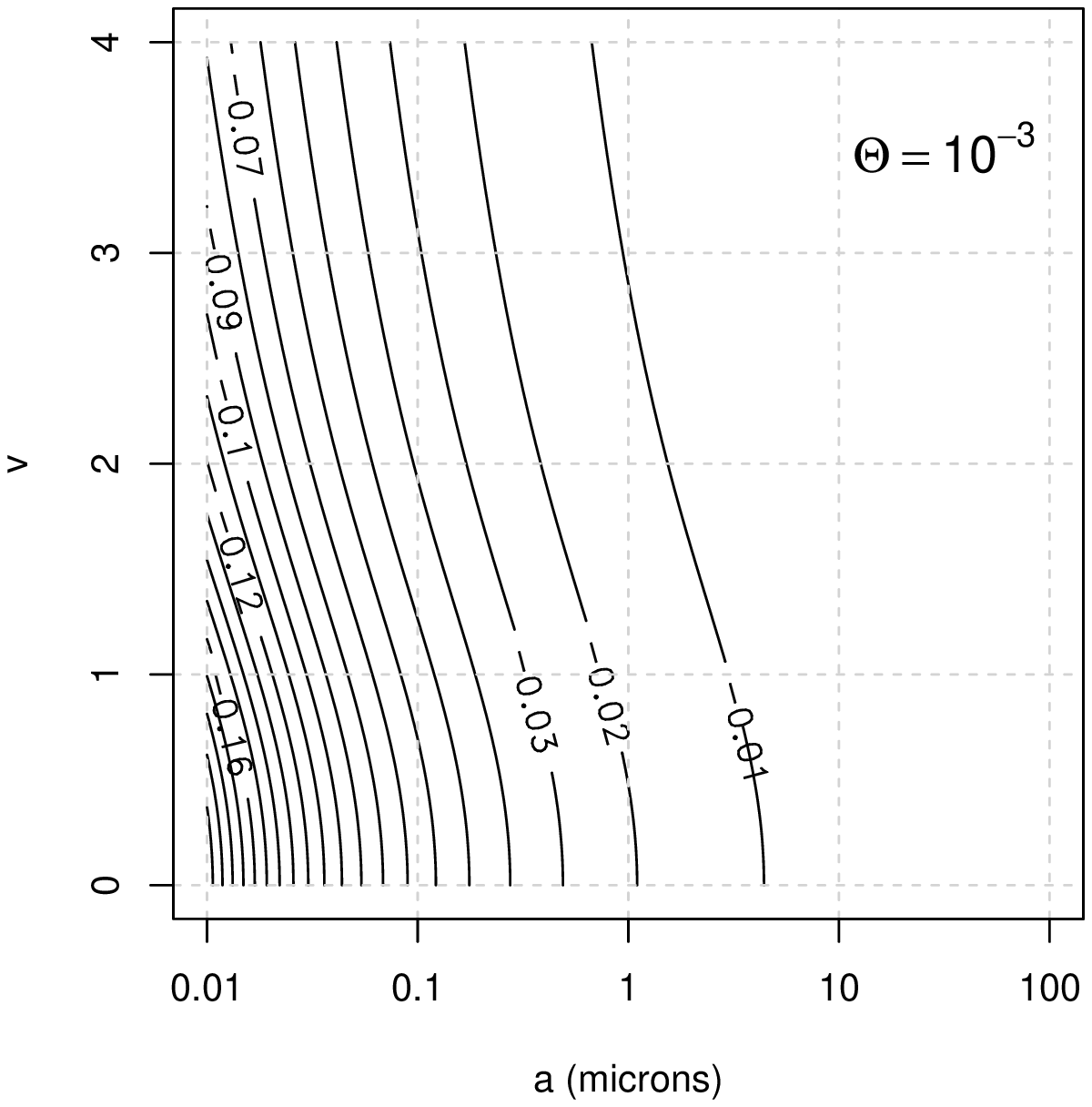}
\caption{\label{fig:skewsA} Contour plots of $f_N$'s skewness for a mid-sized grain in a hydrogenic plasma with 1 eV electrons. \emph{Top:} $\Theta = 1$. \emph{Bottom:} $\Theta = 10^{-3}$.}
\end{figure}

\begin{figure}
\includegraphics[width=0.49\textwidth]{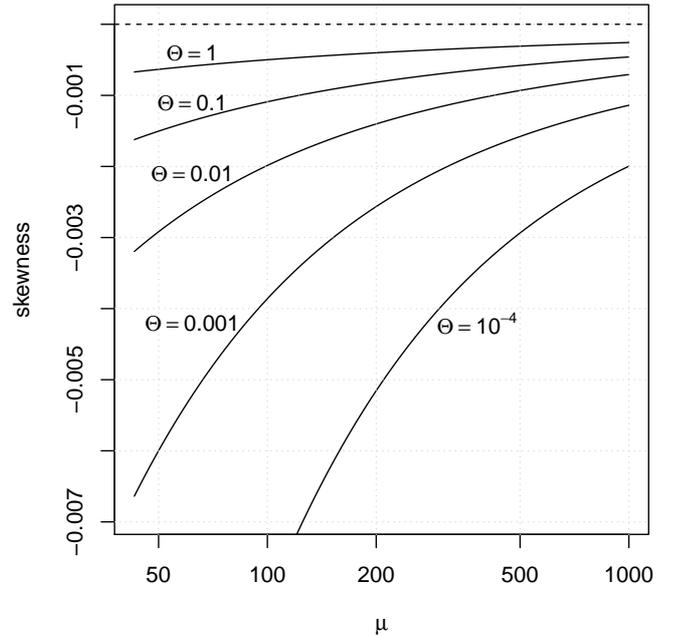}
\caption{\label{fig:skewsmu} $f_N$'s skewness as a function of $\mu$ for a 10 nm grain ($a \lesssim \lambda_D$) in a stationary plasma with 1 eV electrons.}
\end{figure}

We explore this possibility for mid-sized grains first. To assess $f_N$'s skewness, we compute it for a hydrogenic plasma with 1 eV electrons, as a function of the normalized flow velocity $v$ and the sphere's radius $a$ (figure \ref{fig:skewsA}). Numerical experiments reveal that the skewness is less with heavier ions (figure \ref{fig:skewsmu}), so the hydrogenic plasma results we discuss here are a worst-case scenario.

Figure \ref{fig:skewsA} shows decreasing skewness with increasing radius and flow speed. With equal ion and electron temperatures ($\Theta = 1$) skewness is consistently small, except in the limit of vanishing sphere radius, but in that limit the model becomes invalid anyway as $\alpha \gg 1$. With cooler ions, skewness is greater and less affected by flow.

\begin{figure}
\includegraphics[width=0.49\textwidth]{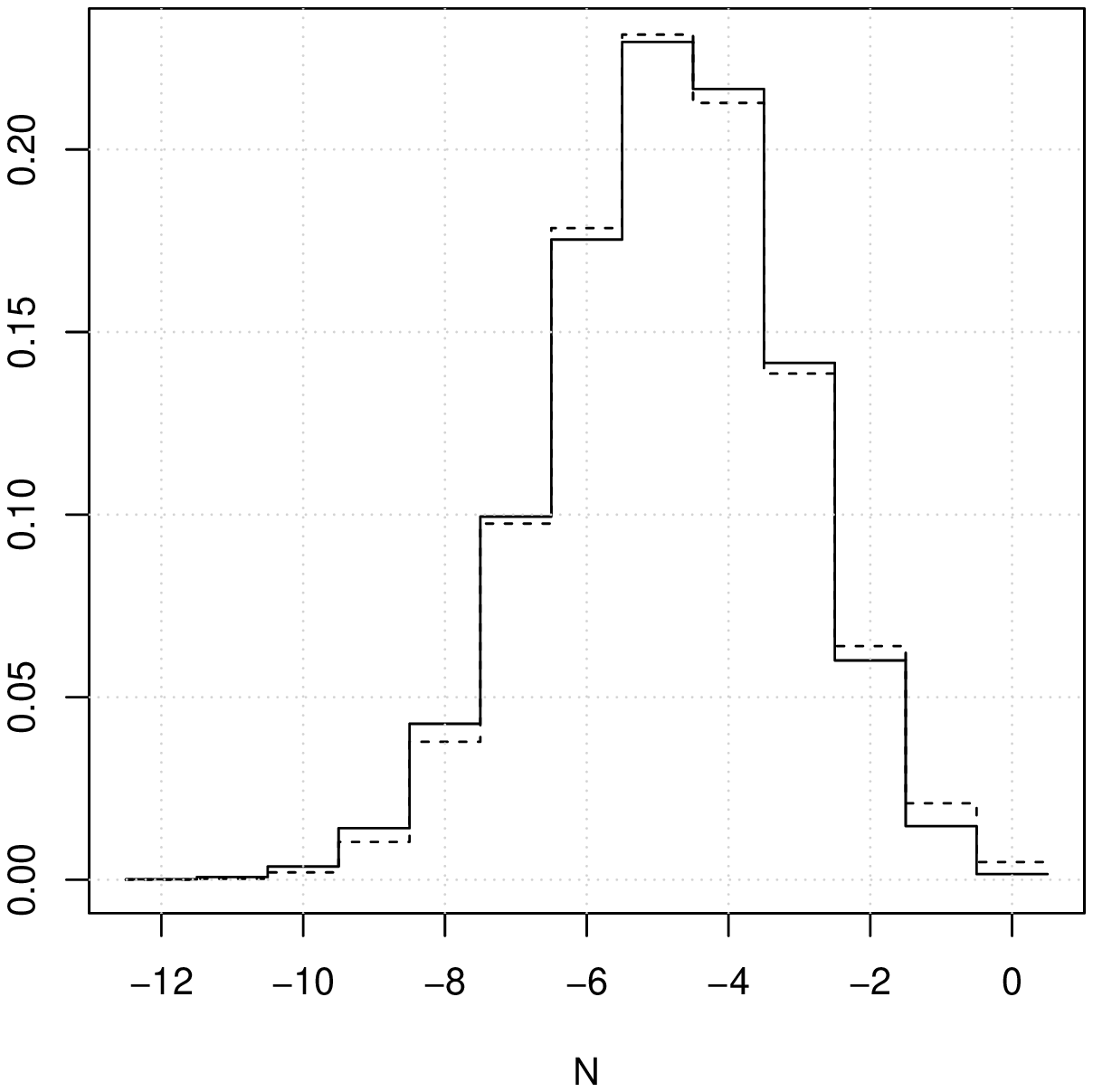}
\caption{\label{fig:fN1} $f_N$ (solid line) and its Gaussian approximation (dashed line) for a 10 nm sphere in a hydrogenic plasma with $a \lesssim \lambda_D$, $v = 0$, $\Theta = 0.001$, and 1 eV electrons.}
\end{figure}

For a 10 nm sphere in a stationary hydrogenic plasma with 1 eV electrons and $\Theta = 0.001$, $f_N$'s skewness is $-0.206$. (This corresponds to the bottom left corner of the lower plot in figure \ref{fig:skewsA}.) This is a non-negligible but nonetheless modest degree of skew, and the Gaussian approximation holds up well (figure \ref{fig:fN1}). That it does so even for these inconvenient parameter values suggests that the Gaussian approximation is robust.

The large grain model's $f_N$ is even less skewed for realistic parameter values. For large grains the skewness depends on the four parameters $\alpha$, $\Theta$, $v$ and $\gamma$, but as it increases only marginally with $\gamma$ over the range $1 \le \gamma \le 3$ we may ignore $\gamma$.

Like the mid-sized grain model, the skewness decreases with increasing $\mu$. Unlike the mid-sized grain model, the large grain model's skewness \emph{decreases} with $\Theta$ (except when $\alpha \sim 1$, $v$ is huge, and $\Theta$ is already very small). This fits our finding above that for cold ions the mid-sized grain model breaks down, while the large grain model improves its self-consistency.

\begin{figure}
\includegraphics[width=0.49\textwidth]{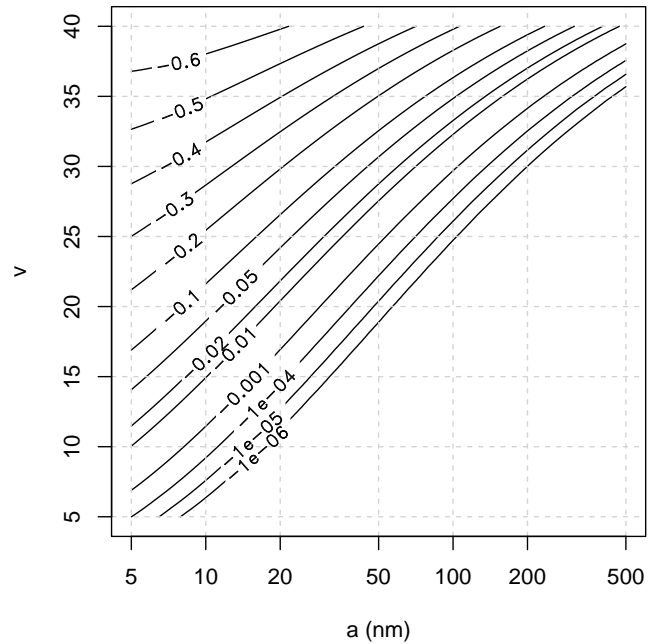}
\caption{\label{fig:skewsB} Contour plot of $f_N$'s skewness for a large grain in a hydrogenic plasma with 1 eV electrons and ions, $v \gg 1$, and $\gamma = 3$.}
\end{figure}

The foregoing means that $f_N$'s skewness is highest for large grains when $\gamma$, $\Theta$, $\alpha$ and $v$ are high. Figure \ref{fig:skewsB} presents numerical calculations of $f_N$'s skewness where $\Theta$ and $\gamma$ take on their highest realistic values (1 and 3 respectively). Even in this most pessimistic case, apprecible skewness is only a risk when $v$ is enormous or the grain's size tends towards the nanometre scale, and is merely a symptom of the large grain model failing as its validity conditions are progressively violated. When those conditions are instead satisfied, the skewness of the predicted charge distribution is negligible, and the large grain model's Gaussian approximation appears to be even more robust than the mid-sized grain model's.

\section{Conclusion}

We have derived equilibrium probability distributions of a spherical grain's charge in a flowing, collisionless plasma, using stochastic models based on the SOML charging theory. It transpires that these distributions are expressible in closed form in terms of exponential and gamma functions. The modal grain charge is proportional to $T_e a$ for large grains, and remains approximately proportional to $T_e a$ for mid-sized grains with small $\alpha$.

When the grain is large enough (and the ions are no hotter than the electrons, which is usually true) Gaussian distributions approximate the exact distributions well, affirming Matsoukas et al.'s demonstration that particle charge ``fluctuations are Gaussian, regardless of the detailed form of the charging currents'' \cite{Matsoukas96}. For mid-sized grains, the Gaussian distribution's variance increases with the normalized flow velocity $v$, with the dependence on $v$ strongest for $v \approx 1$. For large grains there is no $v$ dependence.

One possible use of the Gaussian approximation is estimating the likelihood of various deviations from the mean charge. For example, in a low-pressure argon discharge with $n_e = 10^{16}$ m$^{-1}$ and $T_e = 100 T_i = 4$ eV, a 10 $\mu$m sphere ($a \ll \lambda_D = 149$ $\mu$m) has a 53\% chance of being within 0.1\% of its equilibrium charge, while the same sphere in a tokamak edge deuterium plasma ($n_e = 10^{20}$ m$^{-1}$, $T_e = T_i = 100$ eV, and $a > \lambda_D = 7$ $\mu$m) has a $>$99\% chance of being within 0.1\% of its equilibrium charge.

There are many ways in which this paper's results might be extended. For example, we assume our spheres are conducting yet remain unpolarized in the face of approaching electrons and ions. Physically this is a contradiction in terms, but it simplifies our calculations and has a substantial effect only on tiny spheres ($a \lesssim 10$ nm) in plasmas with cool electrons ($k_B T_e \lesssim 1$ eV) \cite{Matsoukas96}. It might nonetheless be useful to incorporate electrical polarization into our models. Our models might also be generalizable to magnetized plasmas, collisional plasmas, grains out of equilibrium, non-spherical grains, plasmas with non-Maxwellian electron velocity distributions, grains in a sheath, and grains that emit electrons (whether by photoelectric, thermionic, or field emission). As our models stand, however, they should remain applicable to a broad range of close-to-spherical grains in typical flowing plasmas.

\appendix

\section{Intractability of the small sphere case with a flowing plasma}

\label{S:intractability}

For a negatively charged sphere in a singly ionized plasma, the SOML ion current is
\begin{equation}
I_i = 4 \pi a^2 n_0 e \sqrt{\frac{k_B T_i}{2 \pi m_i}} \left( s_1(v) - s_2(v) \frac{e \phi}{k_B T_i} \right)
\end{equation}
which is equation \ref{eq:somlIi} above with $Z = 1$.
(There seems to be a typo, incidentally, in Willis et al.'s presentation \cite{Willis12} of this result. Their equation 13 lacks an exponent of $1/2$ for the parenthetical fraction.) However, a tiny sphere has a non-negligible chance of being positively charged. Under SOML a positively charged sphere has the ion current \cite{Shull78,Willis12}
\begin{equation}
I_i = 4 \pi a^2 n_0 e \sqrt{\frac{k_B T_i}{2 \pi m_i}} \left( s_3(v) - s_4(v) \frac{e \phi}{k_B T_i} \right)
\end{equation}
where the new velocity-dependent auxiliary functions are
\begin{align}
s_3(v) \equiv & \ \sqrt{\pi} \frac{1 + 2v^2}{8v} (\!\erf(v + v_0) + \erf(v - v_0)) \\
& + \frac{1 + v_0 / v}{4} \exp \! \left( -(v - v_0)^2 \right) \notag \\
& + \frac{1 - v_0 / v}{4} \exp \! \left( -(v + v_0)^2 \right) \notag
\end{align}
and
\begin{equation}
s_4(v) \equiv \sqrt{\pi} \frac{\erf(v + v_0) + \erf(v - v_0)}{4v}
\end{equation}
with $v_0 \equiv \sqrt{e \phi / (k_B T_i)}$ as a normalized speed cutoff.
These auxiliary functions are complicated functions of $\phi$ via $v_0$, and this blocks an exact analytic solution for $f_N$ where $N>0$. We can and do ignore this for larger spheres as they're almost always negatively charged, but for spheres small enough to attain a positive charge it is an insuperable obstacle.

Not \emph{all} is lost: one can derive a complete solution for $f_N$ with plain OML, i.e.\ when there is no flow. In this special case, one can solve a recurrence relation for $f_N$ when $N \le 0$, solve another recurrence relation for $f_N$ when $N \ge 0$, and graft the two solutions together with the detailed balance condition $r_1 f_1 = g_0 f_0$. We have not done this here as the mechanics of that calculation are little different to those of section \ref{S:solving}, and Draine \& Sutin have already given an analogous solution (albeit assuming $T_i = T_e$) for small grains and low-temperature plasmas \cite[p.\ 808]{Draine87}.

\section{Unimodality of the charge probability distribution}

\label{S:unimod}

Here is a demonstration that $f_N$ is unimodal in the sense of Medgyessy \cite{Medgyessy72}, i.e.\ that the sequence
\begin{equation}
\label{eq:Medgyessy}
\cdots, f_0 - f_1, f_{-1} - f_0, f_{-2} - f_{-1}, \cdots
\end{equation}
has exactly one change of sign after discarding zero terms. In intuitive terms, this asserts that $f_N$ has only one peak, though that peak may spread across multiple adjacent abscissae with the same maximal ordinate.

As this paper's stochastic model assumes a non-positive charge a priori, $f_N = 0 \ \forall \ N > 0$. Hence the term $f_0 - f_1 = f_0$ in sequence \ref{eq:Medgyessy}, and the terms before it are zero and dispensable. Therefore we need only show that
\begin{equation}
\label{eq:halfseq}
f_0, f_{-1} - f_0, f_{-2} - f_{-1}, \cdots
\end{equation}
has one change of sign after discarding zero terms. We now prove this explicitly for mid-sized grains; the same basic logic applies for large grains.

Consider $f_{N-1}/f_N$, given in equation \ref{eq:rec1}. Because $\alpha > 0$, $\exp(\alpha N)$ strictly increases in $N$. Similarly, because $s_1(v)$, $s_2(v)$, and $\alpha$ are always positive, eq.\ \ref{eq:rec1}'s parenthetical divisor strictly decreases in $N$. As the parenthetical divisor must always be non-negative, it follows that eq.\ \ref{eq:rec1} is strictly increasing in $N$ for $N \le 0$. Because $f_{N-1} / f_N$ strictly increases in $N$ for $N \le 0$, $f_{N-1} / f_N - 1$ and hence $f_{N-1} - f_N$ can change sign at most once for $N \le 0$. From the second term onwards, then, sequence \ref{eq:halfseq} has at most one sign change.

Suppose there were such a sign change. This requires that $f_{N-1} / f_N - 1$ changes sign as $N$ becomes more negative, which means $f_{N-1} / f_N$ must go from being more than 1 to being less than 1, because $f_{N-1} / f_N$ decreases as $N$ becomes more negative. This implies that $f_{-1}/f_0 > 1$, implying $f_{-1} - f_0 > 0$, which in turn implies no sign change in sequence \ref{eq:halfseq}'s first two terms (because $f_0>0$). As such, if sequence \ref{eq:halfseq} has a sign change after the second term, it is the only sign change.

Suppose there were instead \emph{no} sign change after the second term. Then either $f_{N-1}/f_N > 1$ for $N \le 0$, or $f_{N-1}/f_N < 1$ for $N \le 0$. The former is impossible, because it asserts that $f_N$ becomes ever larger as $N \rightarrow -\infty$, which would render $f_N$ unnormalizable. The latter implies $f_{-1} < f_0$, and so a sign change between sequence \ref{eq:halfseq}'s first two terms. Thus, were there no sign change after the sequence's second term, there would have to be a sign change between the first two terms.

The last two paragraphs mean that sequence \ref{eq:halfseq} has exactly one sign change, completing the proof that $f_N$ is unimodal.

\end{document}